\documentclass[graybox]{svmult}

\usepackage{type1cm}        
\usepackage{makeidx}         
\usepackage{graphicx}        
\usepackage{multicol}        
\usepackage[bottom]{footmisc}

\usepackage{newtxtext}       %
\usepackage{newtxmath}       

\newcommand{\be}{\begin{equation}}
\newcommand{\ee}{\end{equation}}

\newcommand{\Sp}{{S_+}}
\newcommand{\Sm}{{S_-}}

\makeindex             

\begin{document}

\title*{Kinetic and macroscopic epidemic models in presence of multiple heterogeneous populations}
\titlerunning{Kinetic and macroscopic epidemic models}
\author{Andrea Medaglia \and Mattia Zanella}
\institute{Andrea Medaglia \at Department of Mathematics "F. Casorati", University of Pavia, Italy \email{andrea.medaglia02@universitadipavia.it  } \and Mattia Zanella \at Department of Mathematics "F. Casorati", University of Pavia, Italy \email{mattia.zanella@unipv.it}}
%
%
\maketitle

\abstract*{We study the impact of contact heterogeneity on epidemic dynamics. A system characterized by multiple susceptible populations is considered. The description of the spread of an infectious disease is obtained through the study of a system of Boltzmann-type equations for the number densities of social contacts of the introduced compartments. A macroscopic system of equations characterizing observable effects of the epidemic is then derived to assess the impact of contact heterogeneity. }

\abstract{We study the impact of contact heterogeneity on epidemic dynamics. A system characterized by multiple susceptible populations is considered. The description of the spread of an infectious disease is obtained through the study of a system of Boltzmann-type equations for the number densities of social contacts of the introduced compartments. A macroscopic system of equations characterizing observable effects of the epidemic is then derived to assess the impact of contact heterogeneity.}



\section{Introduction}
\label{sec:1}

The recent efforts to design effective non-pharmaceutical measures to mitigate the COVID-19 pandemic were based on the link between social activities and the spreading of a respiratory disease \cite{Plos}. Several works in mathematical epidemiology characterized the number of contacts of the population taking into account an additional structure that is maintained for the whole dynamics. A classical example is represented by age-structured populations for which realistic contact matrices have been determined, see e.g. \cite{APZ,H96,Novo}. Nevertheless, recent works highlighted strong changes in contact distribution in the early phases of an epidemic, whose evolution can shape the infection dynamics, see \cite{ZV}. For these reasons, in \cite{DPeTZ} it has been proposed in a simple SIR-type compartmentalization a kinetic model to couple the dynamics of an infectious disease with the contact evolution of a system of agents. At the level of observable quantities, the emerging model is characterized not only by the evolution of densities, like for classical models in compartmental epidemiology, but also by the evolution of the mean number of connections. Interestingly enough, in the present setting models with saturated incidence rates can be easily derived with minimal assumptions \cite{Capasso,XLiu,Zanella_m3as}. Other recent contributions stressed were centered on the effects of the structure of contacts of agents, we mention in this direction the works \cite{DolT,NATSS,Nielsen}. 

In the present contribution, we concentrate on the influence of contact heterogeneity on the dynamics of the disease in presence of multiple susceptible populations. Each susceptible compartment can be characterized by its mean number of connections. This situation is very common when non-pharmaceutical interventions have different impacts on the population \cite{buonomo2020} or in presence of sanitary cordon measures, where a portion of the territory results highly affected by the disease. We mention recent contributions in this direction using mobility data  \cite{bertaglia2021b,DellaMLoy,Dutta2020,loy2021,SBKT}.

The mathematical tools that we consider are based on kinetic theory for large interacting systems \cite{Cer,FPTT,PT13} for which we are able to derive the evolution of observable quantities from microscopic, often unobservable, dynamics. In details, we will show how heterogeneity in the contact structure plays a central role in the evolution of an epidemic. In particular, preliminary results will highlight that, in several regimes of parameters, the asymptotic number of recovered can be unexpectedly high in societies with small contact heterogeneity, compared to the ones with high contact heterogeneity. These results are coherent with the recent findings presented in \cite{BBT}.  

In more details, the contribution is organized as follows: in Section \ref{sec:2} we introduce a kinetic compartmental model of interest and we briefly discuss the contact formation dynamics. A macroscopic system of equations is then derived for the coupled evolution of mass fractions and the mean number of connections. In Section \ref{sec:3} we present some results highlighting the impact of contact heterogeneity in the evolution of the disease. Some conclusions and research are then reported in the last Section.

\section{Interplay between contact distribution and epidemic dynamics}\label{sec:2}

In this section, we introduce a kinetic model to describe the spreading of an infectious disease depending on an additional variable describing the number of social contacts. Coherently with the modeling approach introduced in the recent works \cite{DPTZ}, we subdivide the total population into three main compartments: susceptible, who can contract the disease, infected infectious, who can transmit the disease and recovered, corresponding to formerly infected patients that are not infectious. Furthermore, to mimic the early effects of the epidemic, where the collective compliance to reduce the number of daily contacts is often not accepted, we subdivide the susceptible population into two main categories $S_+, S_-$ in relation to their average number of contacts, $m_{S_+}$ and $m_{S_{-}}$ respectively.  

The contact distribution of the whole population is therefore recovered as 
\[
f(w,t) = f_{S_+}(w,t) + f_{S_-}(w,t) + f_I(w,t) + f_R(w,t), \qquad \int_{\mathbb R_+} f(w,t) dw = 1.
\]
Hence, we obtain the mass fractions of population in each compartment and their momentum of order $\alpha>0$ as 
\[
J(t) = \int_{\mathbb R_+} f_J(w,t)dw, \qquad J(t) m_{\alpha,J}(t) = \int_{\mathbb R_+} w^\alpha f_J(w,t)dw. 
\]
Unambiguously, in the following we will indicate the mean of contact in the compartment $J$, corresponding to $\alpha=1$, by $m_J$. 

\subsection{Formation of the contact distribution}\label{sec:contact}

Coherently with \cite{DPeTZ}, we can define a process of contact formation based on microscopic transitions for the variation of contacts of a single agent. At aggregate level, the evolution of the distributions $f_J$, $J \in \{S_{\pm},I,R\}$, can be obtained through a Boltzmann-type equation.  As shown in the aforementioned work, to obtain an explicit formulation of the large time distribution, it is possible to derive the following Fokker-Planck-type equation 
\be \label{eq:FP}
\dfrac{\partial}{\partial t} f_J(w,t) = \dfrac{\lambda_J}{2} \partial_w \left[ \left(\dfrac{w}{m_J}-1\right)f_J \right] + \dfrac{\sigma^2_J}{2} \partial_w^2 (w f_J(w,t))
\ee
%
with $\lambda_J>0$, $\sigma^2_J>0$ and $m_J>0$ the mean number of contacts. 


The emerging large time contact distributions $f^\infty_J(w)$, $J\in \{S_\pm,I,R\}$ of \eqref{eq:FP} can be explicitly computed and are of Gamma-type \cite{To4} coherently with experimental results in \cite{Plos}. In particular, if $\mu_J = \lambda_J/\sigma_J^2>0$, we have
\begin{equation}\label{eq:finf}
f^\infty_J(w) = \left( \dfrac{\mu_J}{m_J} \right)^{\mu_J} \dfrac{1}{\Gamma(\mu_J)} w^{\mu_J-1}\exp\left\{- \dfrac{\mu_J}{m_J}w \right\},
\end{equation}
whose momenta of order $\alpha$ are 
\be\label{eq:mom_gamma}
\int_{\mathbb R_+} w^\alpha f^\infty_J(w)dw = \left( \dfrac{m_J}{\mu_J} \right)^\alpha \dfrac{\Gamma(\mu_J+\alpha)}{\Gamma(\mu_J)} = c_{\alpha,J} m_J^\alpha, 
\ee
where $c_{J,\alpha} = \left( \dfrac{1}{\mu_J}\right)^\alpha \dfrac{\Gamma(\mu_J+\alpha)}{\Gamma(\mu_J)}$. Since $f^\infty_J$ is a Gamma distribution we also have 
\[
c_{\alpha+1,J} = \dfrac{\mu_J+\alpha}{\mu_J} c_{\alpha,J},
\] 
and $m_{\alpha+1,J} = m_{\alpha,J} \dfrac{\alpha+\mu_J}{\mu_J} m_J$. 
We remark that \eqref{eq:finf} is explicitly dependent on the positive parameter $\mu_J = \lambda_J/\sigma^2_J$ that measures the contact heterogeneity of a population in terms of the variance of the distribution of social contacts. More precisely, small values of $\mu_J$ correspond to a larger heterogeneity of the individuals in terms of social contacts.

\subsection{The kinetic model}

The resulting system of kinetic equations is given by 
\be
\label{eq:kin}
\begin{split}
\partial_t f_\Sp(w,t) &= -K(f_\Sp,f_I)(w,t) + \dfrac{1}{\epsilon}Q_\Sp(f_\Sp)(w,t) \\
\partial_t f_\Sm(w,t) &=  -K(f_\Sm,f_I)(w,t)  +  \dfrac{1}{\epsilon}Q_\Sm(f_\Sm)(w,t) \\
\partial_t f_I(w,t) &= K(f_\Sp+f_\Sm,f_I)(w,t) -\gamma f_I(w,t)+  \dfrac{1}{\epsilon}Q_I(f_I)(w,t) \\
\partial_t f_R(w,t)&=  \gamma f_I(w,t) +  \dfrac{1}{\epsilon}Q_R(f_R)(w,t), 
\end{split}
\ee
where $\epsilon>0$ and $\gamma>0$ is the recovery rate. The infection transmission is taken into account by the operator 
\[
K(g,f_I)(w,t) = g(w,t) \int_{\mathbb R_+} \kappa(w,w_*) f_I(w_*,t)dw_*, \qquad g = f_\Sp,f_\Sm,
\]
with $\kappa(w,w_*)>0$ expressing the dependency of the disease transmission by the number of contacts and such that $\kappa(0,y) = \kappa(x,0) = 0$. In \cite{DPTZ} it has been proposed as possible example  
\[
\kappa(x,y) = \beta x^{\alpha_1} y^{\alpha_2}, \qquad \alpha_1,\alpha_2>0.
\]
The operators $Q_J(f_J)$, $J \in \{S_\pm,I,R\}$ characterize the thermalization of the distributions $f_J(w,t)$ and as discussed in Section \ref{sec:contact} are given by 
\[
Q_J(f_J)(w,t) = \dfrac{\lambda_J}{2} \partial_w \left[ \left(\dfrac{w}{m_J}-1\right)f_J \right] + \dfrac{\sigma^2_J}{2} \partial_w^2 (w f_J(w,t)), 
\] 
that are mass and momentum preserving. 

From now on we will omit time dependency. Integrating both sides of \eqref{eq:kin} we get 
\[
\begin{split}
\dfrac{d\Sp}{dt} &= -\beta m_{\alpha_1,\Sp} m_{\alpha_2,I}(t)\Sp I \\
\dfrac{d\Sm}{dt} &= -\beta m_{\alpha_1,\Sm} m_{\alpha_2,I}(t) \Sm I \\
\dfrac{dI}{dt} &= \beta \left[\Sp m_{\alpha_1,\Sp}+\Sm m_{\alpha_1,\Sm}(t)\right] m_{\alpha_2,I} I- \gamma I \\
\dfrac{dR}{dt} &= \gamma I
\end{split}
\]
that is not closed like classical compartmental modeling since it depends on the evolution of local mean values $m_J(t)$. A possible way to obtain a closed system of equations is obtained by resorting to a limit procedure that is classical in statistical physics. Indeed, for $\epsilon \ll 1$ the distribution functions $f_J(w,t)$ collapse to Gamma-type densities with mass fractions $J(t)$ and local mean values $m_J(t)$. After multiplication by $w$ we get 
\[
\begin{split}
\dfrac{d (\Sp m_{\Sp})}{dt} &= - \beta  m_{\alpha_1+1,\Sp} m_{\alpha_2,I}\Sp I,  \\
\dfrac{d (\Sm m_{\Sm})}{dt} &=- \beta  m_{\alpha_1+1,\Sm} m_{\alpha_2,I}\Sm I, \\
\dfrac{d(I m_I)}{dt} &= \beta \left( m_{\alpha_1+1,\Sp} \Sp + m_{\alpha_1+1,\Sm}\Sm \right)m_{\alpha_2,I}I - \gamma m_I I \\
\dfrac{d (R m_R)}{dt} &= \gamma m_I  I
\end{split}\]
Hence, in view of \eqref{eq:mom_gamma} we obtain the following closed system for the evolution of the mass fractions and mean connections in each compartment
\be \label{eq:S2SIR}
\begin{split} 
\dfrac{d\Sp}{dt} &= -\beta c_{\alpha_1,S_+}c_{\alpha_2,I} m_{\Sp}^{\alpha_1} m_{I}^{\alpha_2}\Sp I, \\
\dfrac{d\Sm}{dt} &= -\beta c_{\alpha_1,S_-}c_{\alpha_2,I} m_{\Sm}^{\alpha_1} m_{I}^{\alpha_2} \Sm I, \\
\dfrac{dI}{dt} &= \beta c_{\alpha_2,I}\left[ c_{\alpha_1,S_+}\Sp m_{\Sp}^{\alpha_1}+ c_{\alpha_1,S_-}\Sm m_{\Sm}^{\alpha_1}\right] m_{I}^{\alpha_2} I(t)- \gamma I(t) \\
\dfrac{dR}{dt} &= \gamma I(t)
\end{split}
\ee
and
\be\label{eq:social_mean}
\begin{split}
\dfrac{dm_{\Sp}}{dt}  &= -\dfrac{\beta\alpha_1}{\mu_{S_+}} c_{\alpha_1,S_+}c_{\alpha_2,I} m_{\Sp}^{\alpha_1+1} m_{I}^{\alpha_2}I  \\
\dfrac{dm_{\Sm}}{dt}  &= -\dfrac{\beta\alpha_1}{\mu_{S_-}} c_{\alpha_1,S_-}c_{\alpha_2,I} m_{\Sm}^{\alpha_1+1} m_{I}^{\alpha_2}I  \\
\dfrac{dm_I}{dt} &= \beta c_{\alpha_2,I}  m_I^{\alpha_2} \left\{c_{\alpha_1,S_+} \Sp m_{\Sp}^{\alpha_1} \left(\dfrac{\alpha_1+\mu_{S_+}}{\mu_{S_+}}m_{\Sp}-m_I\right)  \right. \\
&\quad+\left. c_{\alpha_1,S_-} \Sm m_{\Sm}^{\alpha_1} \left(\dfrac{\alpha_1+\mu_{S_-}}{\mu_{S_-}}m_{\Sm}-m_I\right)\right\}\\
\dfrac{dm_R}{dt} &= \gamma (m_I - m_R)\dfrac{I}{R}.
\end{split}
\ee

We observe that the obtained social SIR model with generalized interaction forces reduces to the one obtained in \cite{DPTZ} in the case of a unique susceptible population with the choice $\alpha_1=\alpha_2=1$. 

\subsection{Saturated incidence rate}

Fixing $m_I(t) = \tilde m_I>0$ from the first two equations in \eqref{eq:social_mean} we get
\[
\begin{split}
\dfrac{d m_{S_{\pm}}}{dt} = -\bar \beta_{\pm}(t) m_{S_{\pm}}^{\alpha_1+1},
\end{split}
\]
being $\bar \beta_{\pm}(t) = \dfrac{\beta \alpha_1}{\mu_{S_{\pm}}} c_{\alpha_1,S_{\pm}}c_{\alpha_2,I}\tilde m_I^{\alpha_2}I(t)$ complemented with the initial condition $m_{S_{\pm}}(0)$. The exact solution of the above equation reads
\[
m_{S_\pm}(t) = \dfrac{m_{S_\pm}(0)}{\left( 1 + \frac{\beta\alpha^2_1}{\mu_{S_{\pm}}} m_{S_\pm}^{\alpha_1}(0) c_{\alpha_1,S_{\pm}}c_{\alpha_2,I} \tilde m_I^{\alpha_2}\int_0^t I(s)ds \right)^{1/\alpha_1}}.
\]
Hence, with the introduced assumption we get the following set of first order macroscopic equations with saturated incidence rate
\be
\begin{split} \label{eq:S2SIR_sat}
\dfrac{dS_+}{dt} &= -\beta c_{\alpha_1,S_+}c_{\alpha_2,I} H_+(t,I(t)) S_+ I,  \\
\dfrac{dS_+}{dt} &= -\beta c_{\alpha_1,S_-}c_{\alpha_2,I} H_-(t,I(t)) S_- I, \\
\dfrac{dI}{dt}   &= \beta c_{\alpha_2,I} (c_{\alpha_1,S_+}H_+(t,I(t))S_+ + c_{\alpha_1,S_-}H_-(t,I(t))S_-) I - \gamma I \\
\dfrac{dR}{dt}   &= \gamma I(t),
\end{split}
\ee
where
\[
H_\pm(t,I(t))=\dfrac{\tilde m_I^{\alpha_2} m_{S_{\pm}}^{\alpha_1}(0)}{1 + \frac{\beta\alpha^2_1}{\mu_{S_{\pm}}} m_{S_\pm}^{\alpha_1}(0) c_{\alpha_1,S_{\pm}}c_{\alpha_2,I} \tilde m_I^{\alpha_2}\int_0^t I(s)ds}
\]
is a generalization of the classical saturated incidence rate.

To understand the influence of contact heterogeneity we  divide the equations for $S_\pm(t)$ in \eqref{eq:S2SIR_sat} by $dR/dt$. Hence, in the limit $t\rightarrow+\infty$ we have
\begin{equation}\label{eq:SpmR}
\dfrac{dS^\infty_\pm}{dR^\infty}=-\xi_\pm \mu_{S_{\pm}} \dfrac{S^\infty_\pm}{1+\xi_\pm R^\infty},
\end{equation}
where $\xi_\pm=\dfrac{\beta c_{\alpha_1,S_\pm} c_{\alpha_2,I} \tilde m_I^{\alpha_2} m_{S_{\pm}}^{\alpha_1}(0) }{\gamma \mu_{S_{\pm}}}>0$ is a given constant. The solutions of \eqref{eq:SpmR} are  
\be \label{eq:SR}
S^\infty_\pm(R^\infty) = S_\pm(0)\bigg( 1 + \xi_\pm R^\infty \bigg)^{-\mu_{S_{\pm}}},
\ee
since $S^\infty_\pm(R^\infty=0)=S_\pm(0)$. 

For large times we have
\[
S^\infty_+ + S^\infty_- + R^\infty = 1.
\]
Taking advantage of the explicit solution \eqref{eq:SR}  we can rewrite the last relation as 
\be \label{eq:Rmu}
1-R^\infty = S_+(0)\bigg( 1 + \xi_+ R^\infty \bigg)^{-\mu_{S_{+}}} + S_-(0)\bigg( 1 + \xi_- R^\infty \bigg)^{-\mu_{S_{-}}},
\ee
whose solution defines the dependence of $R^\infty$ by the introduced contact heterogeneity. 

\section{Numerical results}\label{sec:3}

In this section, we present several numerical experiments for the system \eqref{eq:S2SIR}-\eqref{eq:social_mean} and the system \eqref{eq:S2SIR_sat} with saturated incidence rate. In particular, we focus on the relation between the fraction of the recovered at the equilibrium $R^\infty$ and the coefficients $\mu_{S_{\pm}}$ measuring the heterogeneity of the population of the compartments $S_{\pm}$ in terms of the variance of the contact distribution. More specifically, small values of $\mu_{S_{\pm}}$ correspond to a larger heterogeneity of the individuals with respect to the social contact, since $\mu_{S_{\pm}}=\lambda_{S_{\pm}}/\sigma^2_{S_{\pm}}$. 

\begin{figure}[t]
	\centering
	\includegraphics[scale = 0.39]{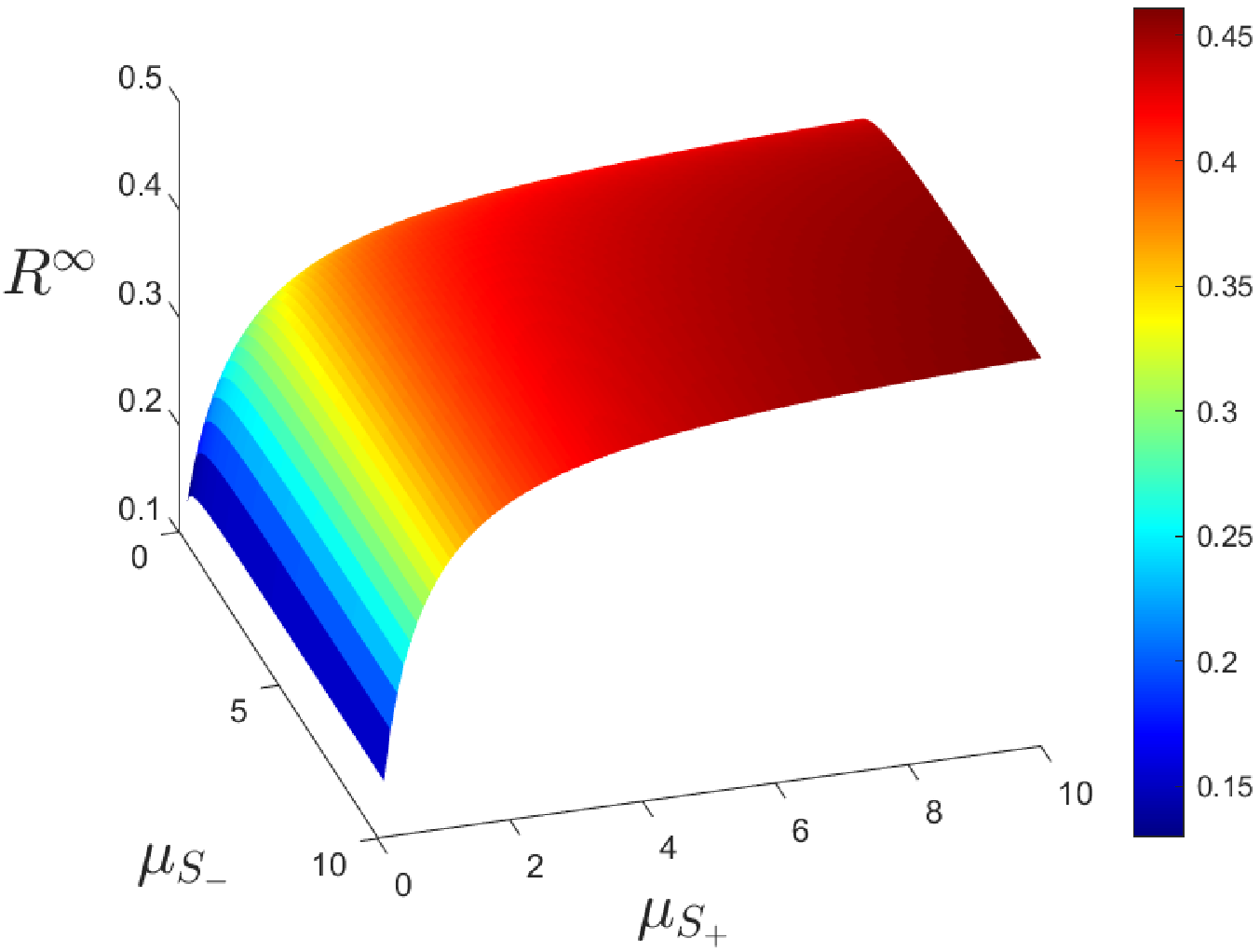}
	\includegraphics[scale = 0.39]{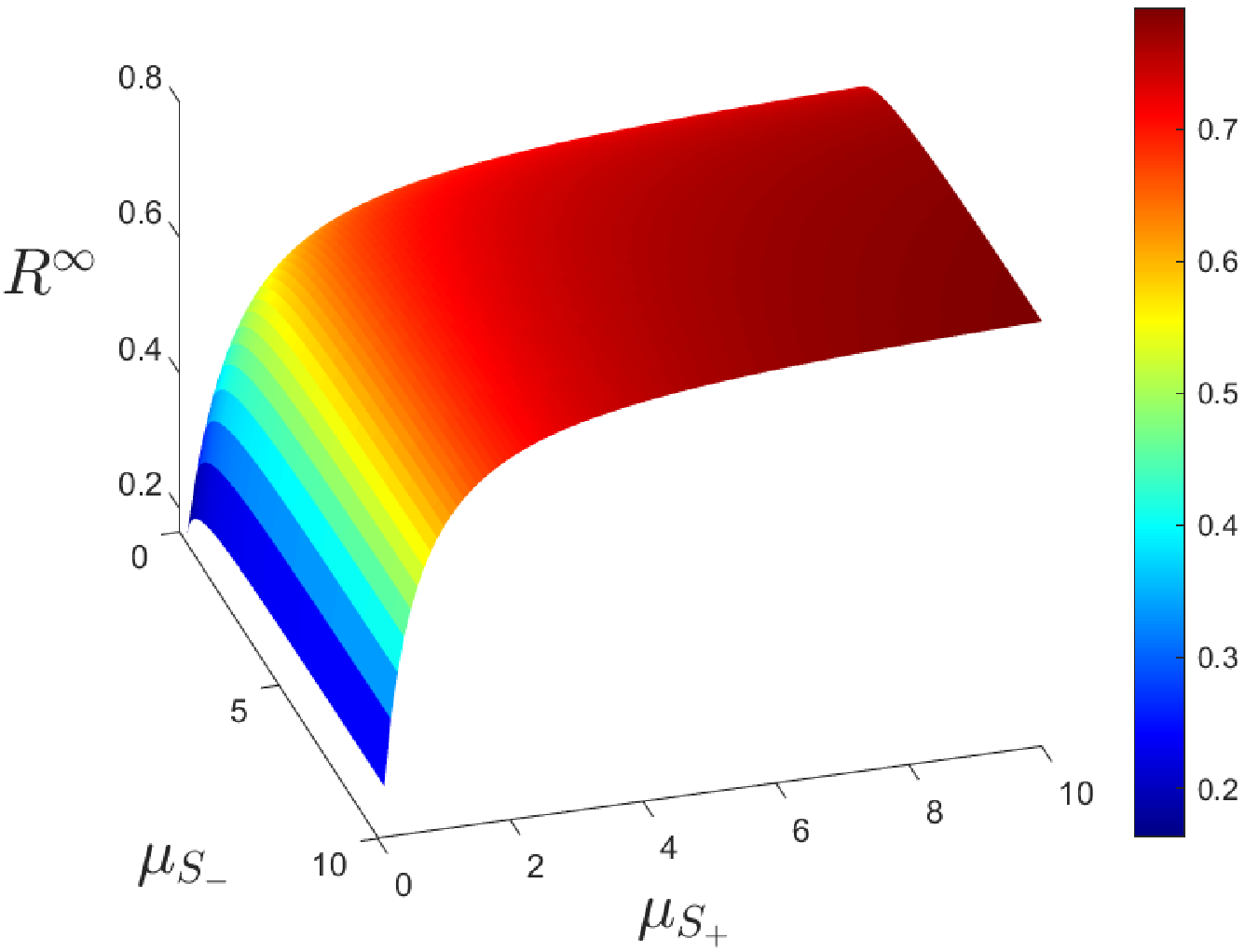}
		\caption{Numerical solution of equation \eqref{eq:Rmu} for $R^\infty$, varying the parameters $\mu_{S_\pm}$. Left: $\mathcal{R}_0=1.25$. Right: $\mathcal{R}_0=2.5$. The initial data are $m_{S_+}(0)=m_R(0)=m_I(0)=\tilde{m}_I=15$, $m_{S_-}(0)=10$, $S_+(0)=0.68$,$S_-(0)=0.28$, $I(0)=R(0)=0.02$, the other parameters of the model $\alpha_1=\alpha_2=1$, $\gamma=0.1$.}
	\label{fig:roots}
\end{figure}


Since $R^\infty$ defined in \eqref{eq:Rmu} depends on several parameters defining the initial set-up of the contact distribution, to understand the influence of the contact heterogeneity, we fix the following values 
\[
m_{S_+}(0)=m_R(0)=m_I(0)=\tilde{m}_I=15, \,m_{S_-}(0)=10,
\]
\[
S_+(0)=0.68,\,S_-(0)=0.28,\,I(0)=R(0)=0.02,
\]
\[
\alpha_1=\alpha_2=1, \,\gamma=0.1.
\]
Therefore, with these choices $R^\infty$ is function of $\mu_{S_+}$ and $\mu_{S_-}$ with a parametric dependence on $\beta$, that is linked to the reproduction number 
\[
\mathcal{R}_0=\dfrac{\beta}{\gamma}m_I(0)\left(S_+(0)m_{S_+}(0)+S_-(0)m_{S_-}(0)\right). 
\]
The relation between the contact structure of the agents and the spreading of the epidemics has been recently studied in literature \cite{BBT,DolT}. In particular, regarding the COVID-$19$ pandemic, it has been pointed out that a smaller heterogeneity could be associated to a larger value of the recovered at the equilibrium. 

Similarly to \cite{BBT}, we consider first the case $\mathcal R_0 = 2.5$. This choice is then compared with  the case of an infectious disease characterized by $\mathcal R_0 = 1.25$. In details, we solve numerically equation \eqref{eq:Rmu} for $R^\infty$, varying $\mu_{S_\pm}$ and taking different values of $\mathcal{R}_0$. As we can observe from Figure \ref{fig:roots}, $R^\infty$ is an increasing function of both the coefficients $\mu_{S_{\pm}}$ regardless of the considered values of the reproduction number $\mathcal{R}_0$. 

A rather different behavior can be observed for the \textit{non-saturated} system \eqref{eq:S2SIR}-\eqref{eq:social_mean}. In particular, $R^\infty$ exhibits in this case a maximum for small $\mu_{S_\pm}$. This means that a higher heterogeneity is linked to a larger value of the recovered at the equilibrium. As a consequence, we note also that different conditions of the heterogeneity of the social contacts could be associated to the same $R^\infty$, despite they have a distinct time evolution.   

\begin{figure}[t]
	\centering
	\includegraphics[scale = 0.39]{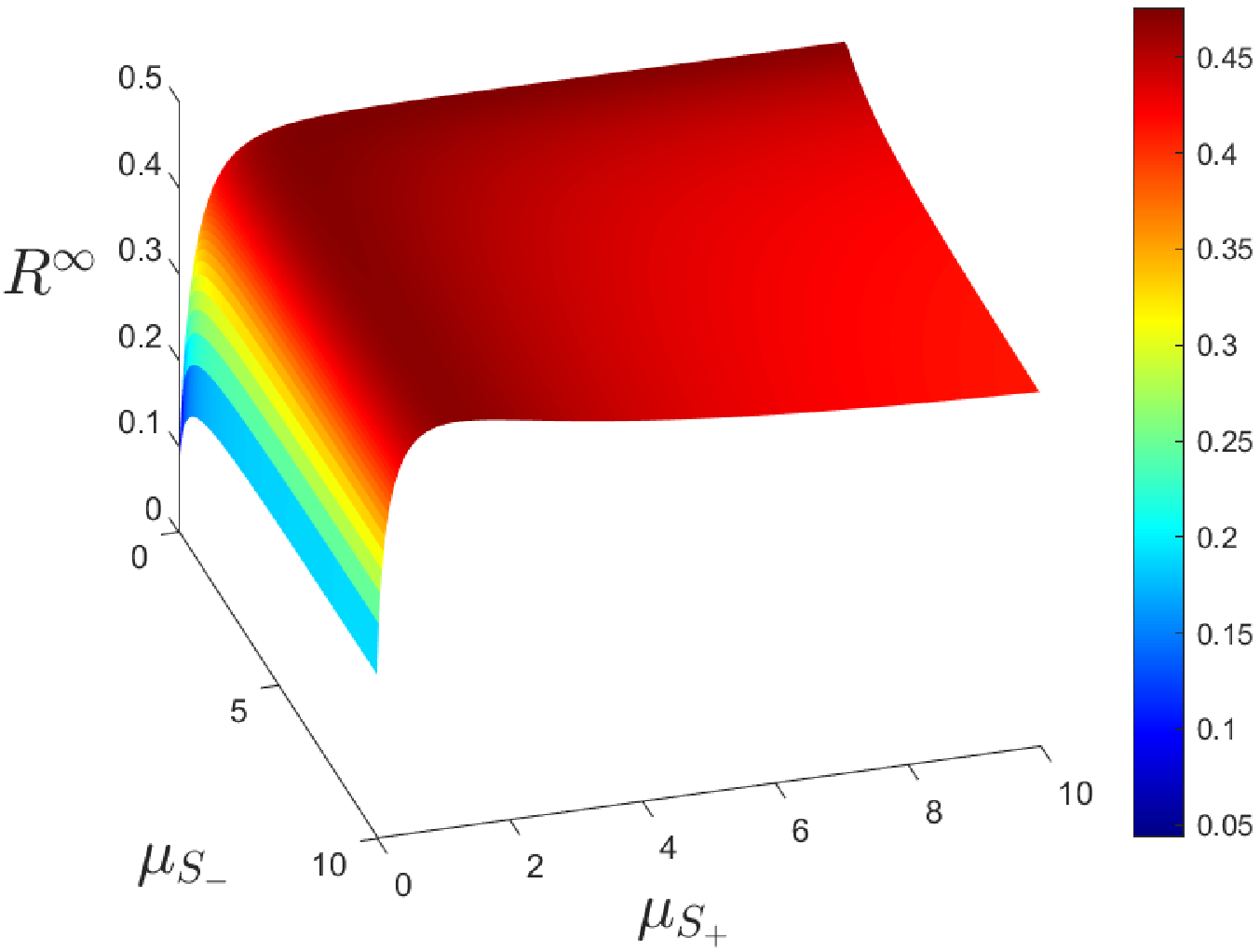}
	\includegraphics[scale = 0.39]{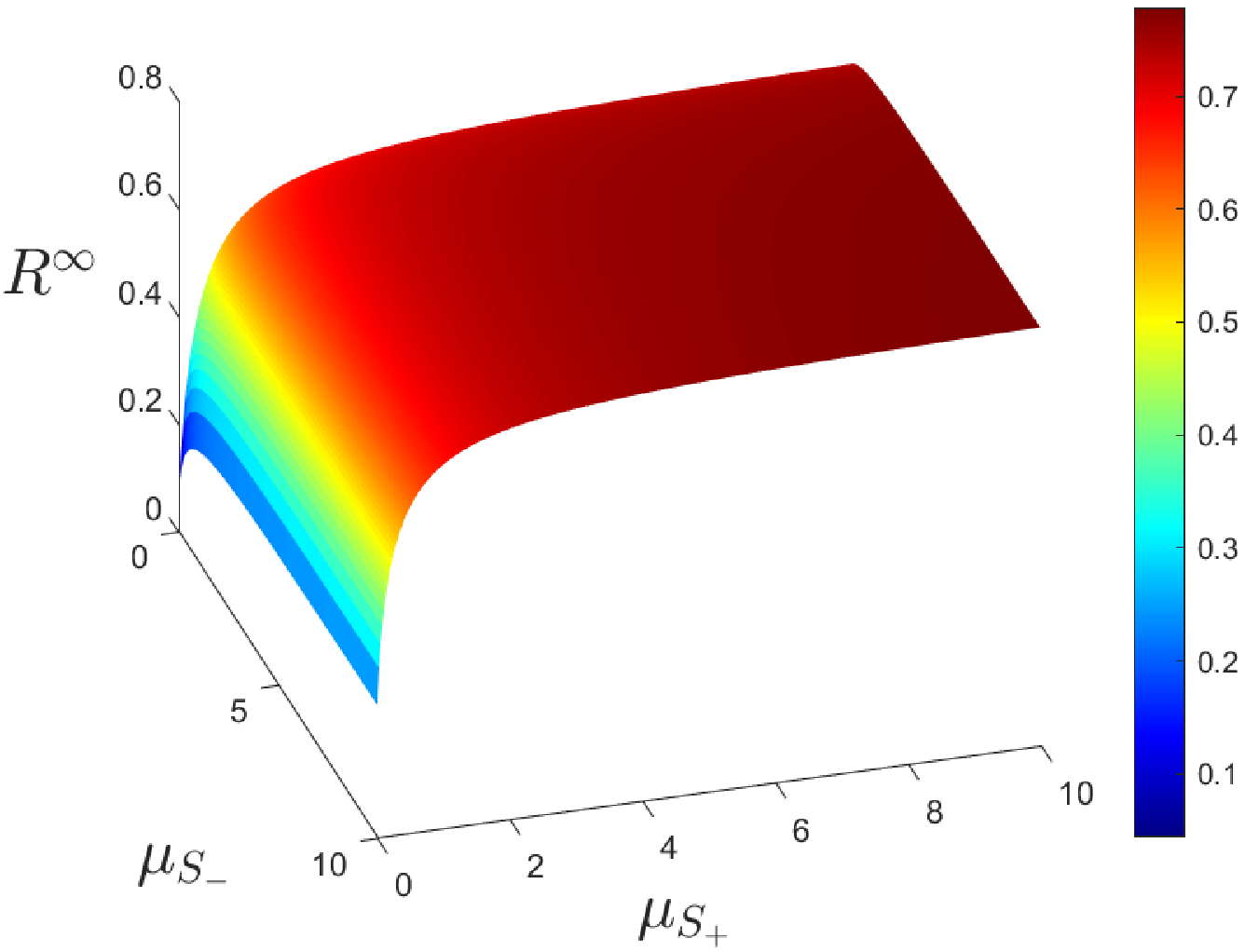}
		\caption{Fraction of recovered $R^\infty$ versus the parameters $\mu_{S_\pm}$, obtained solving the system \eqref{eq:S2SIR}-\eqref{eq:social_mean}. Left: $\mathcal{R}_0=1.25$. Right: $\mathcal{R}_0=2.5$. The initial data are $m_{S_+}(0)=m_R(0)=m_I(0)=\tilde{m}_I=15$, $m_{S_-}(0)=10$, $S_+(0)=0.68$,$S_-(0)=0.28$, $I(0)=R(0)=0.02$, the other parameters of the model $\alpha_1=\alpha_2=1$, $\gamma=0.1$.}
	\label{fig:NonSat}
\end{figure}

\begin{figure}[t]
	\centering
	\includegraphics[scale = 0.45]{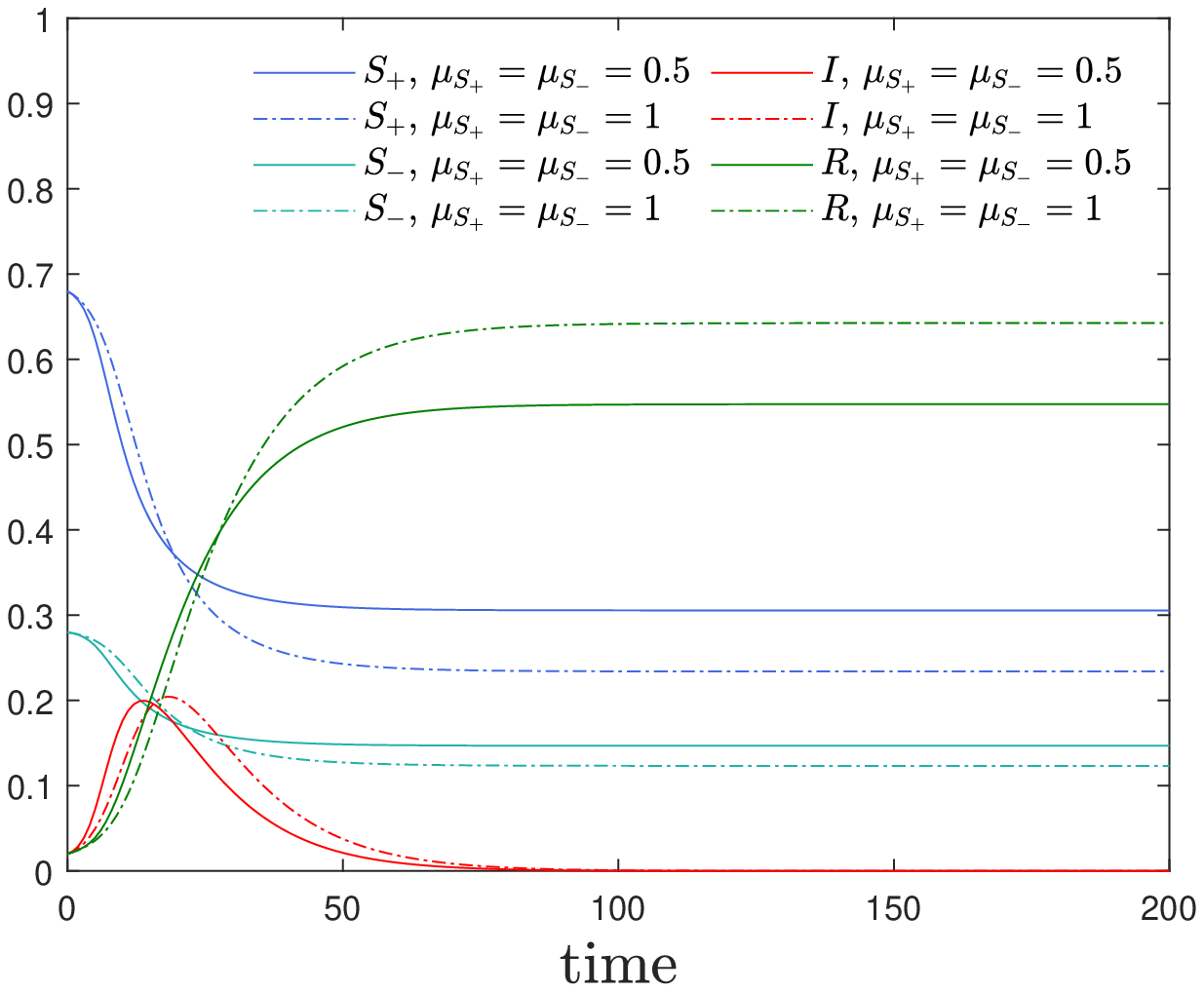}
	\includegraphics[scale = 0.45]{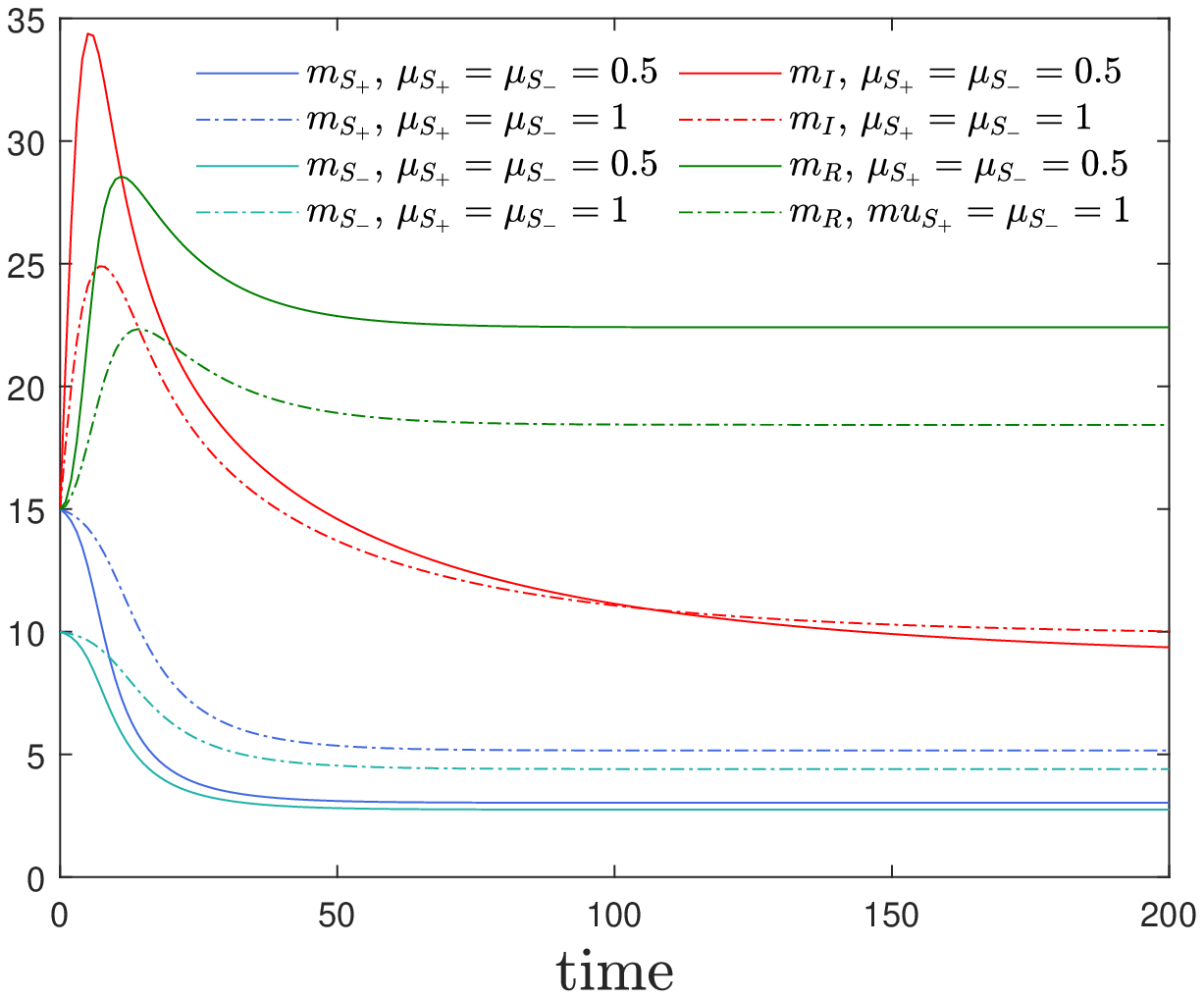}
		\caption{Evolution of the system \eqref{eq:S2SIR}-\eqref{eq:social_mean} with $\mathcal{R}_0=2.5$ for $\mu_{S_+}=\mu_{S_-}=0.5$ (solid lines) and $\mu_{S_+}=\mu_{S_-}=1$ (dashed lines). The initial conditions are $m_{S_+}(0)=m_I(0)=m_R(0)=15$, $m_{S_-}(0)=10$, $S_+(0)=0.68$, $S_-(0)=0.28$ and $I(0)=R(0)=0.02$. The other parameters are $\alpha_1=\alpha_2=1$, $\gamma=0.1$.}
	\label{fig:S2SIR}
\end{figure}

As we can easily observe from the left panel of Figure \ref{fig:NonSat}, $R^\infty$ in the $\mathcal{R}_0=1.25$ scenario has a maximum for high conditions of heterogeneity and then it decreases as the parameters $\mu_{S_\pm}$ increase. On the contrary, the $\mathcal{R}_0=2.5$ case in the right panel of Figure  \ref{fig:NonSat} shows the same trend of the system with saturated incidence rate. In more details, in Figure \ref{fig:S2SIR} we show the time evolution of the system \eqref{eq:S2SIR}-\eqref{eq:social_mean} for $\mathcal{R}_0=2.5$. We clearly see that a decreasing contact heterogeneity is associated to bigger fraction of recovered for large times.    

In the end, we observe that for high values of heterogeneity parameters the model with saturated incidence rate, mimicking non-pharmaceutical protection measures such as a lockdown strategy, exhibits a lower fraction of recovered at the equilibrium than the system \eqref{eq:S2SIR}-\eqref{eq:social_mean}. In particular, for small $\mathcal{R}_0$, a fixed mean of contacts in the infected compartment is able to avoid the maximum for small $\mu_{S_\pm}$ shown in the left panel of Figure \ref{fig:NonSat}. 

\section*{Conclusion and perspectives}

In this short note, we focused our attention on a kinetic compartmental model describing the spread of an infectious disease. The process of contact formation is coupled with the epidemic dynamics. We show that the presence of contact heterogeneity is central for the assessment of the evolution of a disease. The interplay between the process leading to the formation of social contacts and Maxwellian models with multiple interactions studied in \cite{BCG} is currently under deeper investigation. 

\begin{acknowledgement}
This work has been written within the activities of GNFM group of INdAM (National Institute of High Mathematics). The research was partially supported by the Italian Ministry of Education, University and Research (MIUR): Dipartimenti di Eccellenza Program (2018– 2022) - Dept. of Mathematics “F.Casorati”, University of Pavia. 
\end{acknowledgement}

\end{document}